\documentclass{optica-article}

\journal{opticajournal} 

\articletype{Research Article}

\usepackage{lineno}

\usepackage{multirow}
\usepackage{cleveref}
\usepackage{physics}

\usepackage{afterpage}

\usepackage{xcolor}
\definecolor{colour1}{RGB}{204,120,188}
\definecolor{colour2}{RGB}{86,180,233}
\definecolor{colour3}{RGB}{213,94,0}
\definecolor{cbl0}{RGB}{1,115,178}
\definecolor{cbl1}{RGB}{222,143,5}
\definecolor{cbl2}{RGB}{2,158,115}
\definecolor{cbl4}{RGB}{204,120,188}
\definecolor{cbl9}{RGB}{86,180,233}

\crefname{Figure}{Fig.}{Figs.} 

\begin{document}

\title{Coexistence of Entanglement-based Quantum Channels with DWDM Classical Channels over Hollow Core Fibre in a Four Node Quantum Communication Network}

\author{
Marcus~J~Clark\authormark{1,2,*}, 
Obada~Alia\authormark{1,4}, 
Sima~Bahrani\authormark{1}, 
Gregory~T~Jasion\authormark{3}, 
Hesham~Sakr\authormark{3}, 
Periklis~Petropoulos\authormark{3}, 
Francesco~Poletti\authormark{3}, 
George~T~Kanellos\authormark{1,6}, 
John~Rarity\authormark{2}, 
Reza~Nejabati\authormark{1,5}, 
Siddarth~K~Joshi\authormark{2,†}, 
Rui~Wang\authormark{1,‡},  
Dimitra~Simeonidou\authormark{1} 
}   

\address{
\authormark{1}High Performance Networks Group \& School of Electrical, Electronic, and Mechanical Engineering, University of Bristol, Bristol, BS8 1TH, UK\\
\authormark{2}Quantum Engineering Technology Labs \& School of Electrical, Electronic, and Mechanical Engineering, University of Bristol, Bristol, BS8 1FD UK\\
\authormark{3}Optoelectronics Research Centre, University of Southampton, SO17 1BJ, UK\\
\authormark{4}now with Global Technology Applied Research, JPMorganChase, Singapore, 486036\\
\authormark{5}now with Outshift by Cisco, California, United States\\
\authormark{6}now with the National and Kapodistrian University of Athens, Athens 157 72, Greece.
}

\email{
\authormark{*} mj.clark@bristol.ac.uk
\authormark{†}sk.joshi@bristol.ac.uk
\authormark{‡}rui.wang@bristol.ac.uk
} 


\begin{abstract*} 
We experimentally demonstrate the coexistence of three entanglement-based quantum channels with carrier-grade classical optical channels over $11.5$\,km hollow core nested antiresonant nodeless fibre, in a four user quantum network. 
A transmission of 800~Gbps is achieved with four classical channels simultaneously with three quantum channels all operating in the C-band with a separation of $1.2$\,nm, with aggregated coexistence power of $-3$\,dBm. 
We established quantum key distribution in the four-node full-mesh quantum network with Bell state fidelity of up to $90.0\pm0.8$\%. 
The secret key rate for all the links in the network are passively preserved over $55$\,hours of experimental time.
\end{abstract*}

\section{Introduction}
\label{sec:intro}

The landscape of quantum communication networks has evolved from point-to-point links to multi-user networks with practical applications~\cite{Cao2022TheQinternet, Mehic2020QuantumPerspective, chapuran2009optical}. 
A key advancement in this field is the integration of entanglement-based networks, as demonstrated through passive and active architectures~\cite{wengerowsky2018entanglement, joshi2020trusted, wang2022dynamic}. 
The entanglement resource enhances the security of quantum key distribution~\cite{ekert1991quantum, bennett1992quantum, zapatero2023advances} and facilitates the seamless coordination of information across interconnected nodes, laying the foundation for the envisioned global quantum network~\cite{wehner2018quantum, Sharma2021}. 
Furthermore, the integration of classical channels with quantum channels over the existing telecommunication infrastructure emerges as a critical consideration for the cost-effective deployment and commercialization of quantum networks~\cite{Dou:24}.

Undertaking this integration presents a great challenge due to the substantial power difference between classical channels and quantum channels.
The classical channels introduce additional noise during propagation along the fibre, primarily stemming from optical non-linear effects such as Raman scattering~\cite{subacius2005backscattering, da2014impact, Eraerds_2010}, and crosstalk of classical channels. 
Coexistence of entanglement-based quantum channels with classical channels in the same optical bands, e.g. C-band, over short fibre distances and different optical bands, e.g. O-band for the quantum channels and C-band for the classical channel, have been demonstrated~\cite{fan2023energy, thomas2022entanglement}. 
However, in these demonstrations the additional noise due to the classical communications led to an increase in the Quantum Bit Error Rate (QBER), causing a drastic reduction in performance, even to the point where Quantum Key Distribution (QKD) fails.

Ensuring high isolation (>110 dB) \cite{peters2009dense} between the quantum and classical channels is crucial to minimise the impact of the crosstalk when aggregating and separating quantum and classical channels.
Typically, this isolation is achieved through a sharp filter or the cascading of multiple filters, thereby reducing degradation to the quality of the quantum signal. 
Further, the classical signals are bright enough that non-linear optical processes occur within standard solid core fibre. These generate photons at other wavelengths and pollute the quantum channels.
This could be mitigated by careful power selection of coexisted classical channels and their channel allocation with respect of quantum channels. 
Other potential solutions include using either free-space optical communication systems or adopting a novel type of fibre with a low-nonlinearity.

Recently, a revolutionary new type of fibre technology was developed based on antiresonant structures, called Antiresonant Hollow Core Fibres (AR-HCF)~\cite{poletti2014nested}. 
This new fibre addressed the challenge posed by non-linear effects during transmission, by guiding light through the hollow core using anti-resonance effect within glass membranes, surpassing the performance of conventional glass core fibres \cite{gao2018hollow}. 
The evolution of Hollow Core fibres (HCFs) has witnessed remarkable advancements; importantly the innovative design of Hollow Core Nested Antiresonant Nodeless fibre (HC-NANFs) \cite{poletti2014nested} which preserve polarisation, and  lowering the loss to less than 0.11dB/km \cite{Chen:24} in a separate demonstration. 
These developments position HC-NANF as a promising candidate for high-performance applications not only in modern communication systems \cite{nespola2021transmission, borzycki2023hollow}, but also to facilitate the simultaneous transmission of classical channels at high optical powers alongside quantum channels~\cite{honz2023first, Alia2022} by nearly eliminating Raman scattering challenges.

To allow polarisation encoding of the quantum signals, the fibre must maintain the polarisation within standard narrow-band communication channels, which normally adopt Dense Wavelength-Division Multiplexing (DWDM), either in 50 GHz or 100 GHz spacing  ~\cite{Trenti2024HighFidelityPolarisation}.
Anti-resonant Hollow Core Fibres (AR-HCF) have been shown to support polarisation entanglement~\cite{Chen:21, Trenti2024HighFidelityPolarisation} and time-bin~\cite{Antesberger2024TelecomEntanglement} entanglement distribution links as well as coexistence between telecommunications and QKD~\cite{Alia2022}.
However, to our best knowledge, using HCF in entanglement distribution network to simultaneously distribute different entanglement pairs to different users has not been demonstrated.

In this paper,  we demonstrate an entanglement distribution network supporting the coexistence of up to three entanglement-based quantum channels with four 200~Gbps 16-QAM carrier-grade classical optical channels at a high total coexistence power of -3~dBm over an 11.5~km HCF using 100 GHz DWDM technique. 
The Bell state fidelity of all the entanglement links shows $>88.8$\% in a 3-user configuration and $>80$\% in a 4-user configuration. 
We used the five-ring Hollow Core Nested Anti-resonant Nodeless fibre presented in \cite{nespola2021ultra}, a type of AR-HCF. 
In this work, we only refer to this fibre as HCF.

\section{Experimental System Setup}\label{sec:Testbed}

\label{sec:testbed}
\begin{figure*}[!t]
    \centering
    \includegraphics[width=130mm]{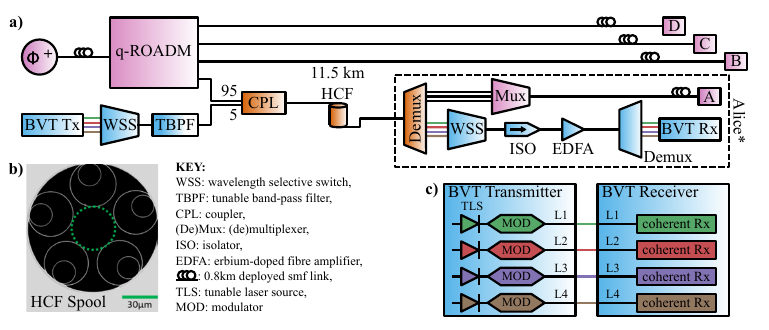} 
    \caption{
    Experimental testbed for quantum-classical coexistence in hollow core fibre.
    \textbf{a)} shows the setup for the coexistence of four classical channels and three entanglement-based quantum channels over 11.5~km HCF in a four-node quantum network.
    \textcolor{colour1}{Purple} components contain only quantum light, \textcolor{colour2}{blue} components contain only classical light, and \textcolor{colour3}{orange} components contain both quantum and classical light.
    \textbf{b)} shows a scanning electron micrograph image of the Hollow Core fibre (HCF) cross-section~\cite{nespola2021ultra}.
    \textbf{c)} shows the classical communications system, consisting of a bandwidth-variable transponder (BVT) transmitter and receiver pair.
    }
    \label{fig:testbed}
\end{figure*}

The HCF-based network testbed used for coexisting the classical and quantum light is shown in \Cref{fig:testbed} \textbf{a)}.
There are three main sections to the physical layer of the network, specifically; where only quantum light is propagated, with components marked in \textcolor{cbl4}{purple}, where only classical light is propagated, with components in \textcolor{cbl9}{blue}, and where classical and quantum light are coexisted, with components noted in \textcolor{cbl1}{orange}.

\subsection{The quantum network}

This testbed uses a central entangled photon source,  providing photon pairs entangled in both the polarisation and time-energy degrees of freedom. 
It enables distribution of a photon in an entangled pair to a network user and the corresponding entangled photon to another network user.
The time-energy degree of freedom is used as a resource, slicing the output spectrum up into discrete wavelength channels following DWDM standard, allowing simultaneous transmission of many entanglement channels at different wavelengths. 
Further details of the source, refer to ~\cite{clark2024quantum}.
This produces an entangled state of,
\begin{equation}
    \ket{\Phi^{+}} = \frac{1}{\sqrt{2}} \left( \ket{HH} + \ket{VV} \right).
\end{equation}

\begin{figure}[t!]
\centering
\vspace{-15pt}
\includegraphics[width=8.5cm]{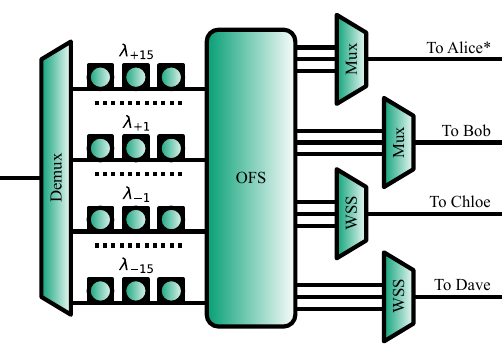}
\caption{
The quantum-enabled reconfigurable optical add-drop multiplexer (q-ROADM) used in the testbed shown in \Cref{fig:testbed} \textbf{a)}.
Here (De)Mux is a (de)multiplexer, 
OFS is an optical fibre switch, 
WSS is a wavelength selective switch,
and FPC is a fibre polarisation controller.
The FPCs are labelled $\lambda_{15}$ to $\lambda_{-15}$ which corresponds to the wavelength pairs, centred where $\lambda_0$ is at ITU channel 34, and $\pm i$ denotes the distance in $100$\,GHz ITU channels from the central channel.
}
\label{fig:q-ROADM}
\end{figure}

To distribute the entanglement resource in this star-network, a quantum-enabled reconfigurable optical add-drop multiplexer (q-ROADM) is employed~\cite{Nejabati2021qroadm}, as shown in \Cref{fig:q-ROADM}.
This system consists of dense-wavelength division multiplexing (DWDM) components, optical switches, and technology to compensate for errors in the degree of freedom where quantum information is encoded. 
In this case, fibre polarisation controllers (FPCs) are used to set the correct polarisation measurement basis across different users in the network.
The DWDM technology allows for the broad spectrum to be separated into $100$\,GHz wide ITU channels, where each ITU channel is entangled to only one other channel.
These are then distributed in pairs where one half of a pair is distributed to one user, and the other half is distributed to another user, allowing the two users to share entanglement.
In this construction, the two users can then recover quantum information by performing correlation measurements in the polarisation degree of freedom. 
In the experiment, we pump the source with a 775.06 nm laser, thus leading to photon pairs generated via SPDC process centred at 1550.12nm. 
This corresponds to channel 34 according to ITU-T 100 GHz standard, where we denote this central wavelength as $\lambda_0$. 
Therefore, $\pm i$ denotes the distance in $100$\,GHz ITU channels from the central channel 34, and photons in $\lambda_{+i}$ are entangled with photons in $\lambda_{-i}$.

This construction allows the q-ROADM to simultaneously share a set of wavelengths to each user, allowing for this simple star-network physical topology to produce a full-connected entanglement mesh across the network.
Such entanglement networks minimises the resources required on the physical layer~\cite{wengerowsky2018entanglement}.
This is achieved through the use of the single central source of quantum states, through the use of both active and passive optical networking technology, instead of constructions where each quantum link requires a transmitter and receiver pair, along with a dedicated optical fibre, for each quantum link.

To compensate for a polarisation definition rotation through optical fibre, a reference light source used as a polarisation definition is integrated inside the entanglement source, allowing all users to match a single definition of polarisation throughout the fibre network.
Since we have four users in this experiment (Alice, Bob, Chloe, and Dave), 12 entangled DWDM-based channels (6 pairs) are required to have a full mesh network -- one pair per link.
Specifically: Alice receives $\lambda_{-15}$, $\lambda_{-14}$, and $\lambda_{-13}$; Bob receives $\lambda_{+15}$, $\lambda_{-12}$, and $\lambda_{-11}$; Chloe receives $\lambda_{+14}$, $\lambda_{+12}$, and $\lambda_{-10}$; and Dave receives $\lambda_{+13}$, $\lambda_{+11}$, and $\lambda_{+10}$,

A user connected to the q-ROADM via fibre will measure polarisation with a polarisation analysis module (PAM).
This device consists of polarisation optics that allow the users to measure both the horizontal-vertical (H-V) polarisation basis and the diagonal-anti-diagonal (D-A) basis with a passive basis selection.
The PAM combines the H and A polarisation measurements onto a common single photon detector (SPD) and the V and D polarisation measurements into another SPD.
This setup reduces the number of SPDs required for a single PAM from 4 to 2, with an increased noise level seen on each measurement basis.
For more information, please refer to Clark, 2024 \cite{clark2024quantum}.

The experiment was conducted across two separate laboratories, the entanglement source, the PAMs, and the SPDs are housed in one building whereas the q-ROADM is housed in a separate building, interconnected through $0.8$\,km deployed telecommunications grade fibre links. 
This was due to equipment limitations, and is not a requirement or feature for this design.
Due to the excellent polarisation stability of these deployed fibre links, no fast adjustments of the polarisation compensation performed by the q-ROADM are required during the experiment.
Three users, Bob, Chloe, and Dave, are directly connected to the q-ROADM through three deployed fibre links, with the first user, Alice*, connected to the q-ROADM via a classical-quantum coexistence fibre link.

\subsection{The classical communications}\label{subsec:ClassicalComms}

The classical communications in the system use an optical DWDM platform with bandwidth-variable transponders (BVTs) as shown in \Cref{fig:testbed}~\textbf{c)}. 
A BVT transmitter is comprised of a tunable laser source (TLS) and a quadrature amplitude modulator (MOD), and a BVT receiver is a coherent receiver.
The unit includes four BVT ports, each being reconfigurable to coherent 100~Gbps (PM-QPSK), 150~Gbps (8-QAM) or 200~Gbps (16-QAM) and can be tuned to any of the 100 wavelengths in the C-band included in the ITU-T grid with 50~GHz offset. 
Adaptable soft-decision forward error correction (SD-FEC) is applied in the transponders to enable error-free communication. 
The BVT receiver enables coherent detection by using a local oscillator for each receiver.
The four classical channels are filtered and combined into a single fibre through a wavelength selective switch (WSS) while any residual out-band noise is further filtered with a tunable bandpass filter (TBPF), as shown in the blue part of \Cref{fig:testbed}.

The BVT modules in this experiment have a maximum output power of $9$\,dBm.
This makes the total maximum power of the 4 channels $15$\,dBm.
After the WSS, TBPF, and the CPL, as shown in \Cref{fig:testbed} \textbf{a)}, the power of each channel was $-9$\,dBm, and in total was a coexistence starting power of $-3$\,dBm.

\subsection{The hollow core fibre}

The HCF used in this demonstration is a combination of three strands, with lengths of $6.2$\,km, $2.6$\,km, and $2.7$\,km, spliced together to produce a single HCF link of $11.5$\,km with no intermediate silica core fibre.
All end-faces of the HCF strands have the same structure and so are compatible for splicing, and an example scanning electron microscopes (SEM) image is shown in \Cref{fig:testbed}~\textbf{b)}. 
Core size, average inter-tube gap, and the membrane thickness for both outer and inner tubes make the fibres operating in the fundamental transmission window at 1550~nm, where details about the fibre parameters can be found in Nespola et al.~\cite{nespola2021ultra}.
The two remaining ends of the combined fibre spool are then spliced to SMF fibre, for compatibility with the fibre-optic components in the testbed.

The loss of the three fibres is $0.98$\,dB/km, $0.85$\,dB/km and $0.95$\,dB/km, respectively, and is spectrally flat across the C-band. 
The total loss of the full 11.5km span is $14.46$\,dB, of which $10.90$\,dB come from fibre propagation. 
The remaining 3.56~dB come from the two HCF-HCF splices, the two SMF-HCF end-splices including mode field adapters (MFAs) and from the connector losses.

The material properties, without a solid silica core, of the HCF lead to ultra-low optical nonlinearity, allowing the transmission of classical channels at a high power with scattering of light, creating out-of-band noise typically expected in SMF.
This then allows multiple channels within the ITU channel grid to have significantly different optical powers, without nonlinear scattered light increasing the noise in the lower power channel, such as a quantum channel.

\subsection{Coexistence}

All the BVT ports were configured with the 16-QAM modulation for a maximum capacity of 200~Gbps per channel resulting in 800~Gbps of transmission overall, as in \Cref{subsec:ClassicalComms}. 
As shown in \Cref{fig:testbed}~\textbf{c)}, the four coherent output ports of the BVT are multiplexed using a wavelength selective switch for a total throughput of 800~Gbps. 
A WSS is used as a multiplexer and a band pass filter to couple the classical channels into a single fibre and provide a 30~dB isolation from the out-band BVT noise. 
The WSS combined output is connected to the input port of a tunable band pass filter (TBPF) with 60~dB of isolation and extremely sharp filter edges to suppress the noise generated by the classical channels. 
The classical channels are then coupled to the quantum channels through a 95:5~coupler (CPL), $95$\% ports for the quantum channels and the $5$\% port for the classical channels, in a co-propagation coexisting configuration and then propagates through the 11.5~km HCF.

\begin{figure*}[!t]
    \centering
    \includegraphics[width=130mm]{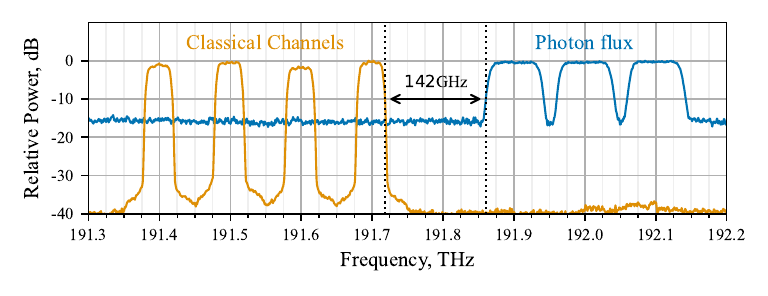}
    \caption{
    The spectral distribution of channels in the coexisted HCF.
    \textcolor{cbl0}{Blue} shows the distribution of Quantum Flux and \textcolor{cbl1}{Orange} shows the distribution of Classical Communications channels.
    The black arrow shows the spectral gap between the 10dB bandwidth edge of the classical signal at $191.7$\,THz and the quantum signal at $191.9$\,THz, with a width of $142$\,GHz.
    The y-axis gives a representative power, relative to the maximum of the given signal type.
    Classical light was measured using the given channel parameters Tab.~\ref{tab:parameters}.
    Photon flux spectrum is measure using ASE noise through exemplar wavelength multiplexing technology.
    }
    \label{fig:SpectralDistribution}
\end{figure*}

After the coexistence, the output is connected into a passive WDM demultiplexer (Demux) that separates the classical and quantum channels based on their wavelengths. 
Next, the quantum channels are filtered and combined using a passive WDM multiplexer (Mux) to eliminate classical channels crosstalk before being transmitted over 0.8~km of SMF to the PAM for user Alice*. 
The output of the four classical ports in Demux are combined using a second WSS and then directed to an optical isolator to prevent light from the tunable laser used by the BVT Rx as a local oscillator from interfering with the measurements of quantum channels. 
It also prevents the Amplified Spontaneous Emission (ASE) noise generated by the Erbium-Doped fibre Amplifier (EDFA) which is used to set the classical signals to the suitable detectable power levels before demultiplexing using a Demux.

Following the experimental and theoretical analysis for optimal wavelength assignment in hybrid quantum-classical networks~\cite{bahrani2019resource, tessinari2021towards}, and based on the fact that the four classical channels are located in the C-band at ITU-T channels 14, 15, 16, and 17 ($1566.31$\,nm - $1563.86$\,nm), as shown in Fig.~\ref{fig:SpectralDistribution}. 
The quantum channels for user Alice are placed at ITU-T channels 19, 20, and 21 (1562.23 - 1560.61~nm) with 142~GHz spacing (1.16~nm) between the 10~dB points of the quantum and classical channels, as shown in Fig.~\ref{fig:SpectralDistribution}. 
This assignment positions the quantum channels at the dip and in the anti-Stokes region of the Raman spectrum of the classical channels, minimising the non-linear effects of the classical channels. 
A summary of coexistence channel parameters is shown in \Cref{tab:parameters}.

\begin{table}[!t] 
\centering
\caption{Parameters for Co-existence Link}
\vspace{-0.3cm}
\label{tab:parameters}
\begin{tabular}{cc}
\hline
\hline \\[-15pt]
\multicolumn{1}{c}{Parameters}  &   \multicolumn{1}{c}{Value}       \\
\hline \\[-15pt]
\multicolumn{2}{c}{\textit{Classical Channels}} \\[-5pt]
Number of Channels  &   4   \\[-4pt]

\begin{tabular}[c]{@{}l@{}} \ \ \ Frequencies \end{tabular}         & 
\begin{tabular}[c]{@{}l@{}@{}l@{}}191.40 THz, 191.50 THz,\\[-5pt] 191.60 THz, 191.70 THz, \end{tabular} \\ [-4pt]

Grid Spacing        &   100 GHz      \\[-4pt]
Modulation Format   &   16-QAM     \\[-4pt]
\begin{tabular}[c]{@{}l@{}}Optical Signal-to-Noise\\  \ \ \ \ \ Ratio (OSNR)\end{tabular} & 20 dB \\[-4pt]
Transmission rate per channel & 200 Gbps      \\[-4pt]
Total transmission rate      &   800 Gbps    \\[-4pt]
Pre-FEC Level       &   15\%        \\[-4pt]
Detector sensitivity*& -26 dBm      \\[3pt]

\hline \\[-15pt]
\multicolumn{2}{c}{\textit{Alice Quantum Channels}} \\[-5pt]
Number of Channels  &   3   \\[-4pt]

\begin{tabular}[c]{@{}l@{}} \ \ \ Frequencies \end{tabular}         & 
\begin{tabular}[c]{@{}l@{}@{}l@{}} 191.90 THz,192.00 THz, 192.10 THz, \end{tabular} \\ [3pt]
\hline
\hline \\[-15pt]
\multicolumn{2}{l}{\textit{*Corresponding to 16-QAM Modulation @200 Gbps.}}  \\
\end{tabular}
\end{table}

\subsection{Raman Scattering}\label{subsec:Raman}

To examine the expected noise at a quantum measurement module, due to Raman scattering phenomena, we must calculate the expected Raman scattered light produced from a classical signal $c_n$, at the wavelength $\lambda_{c_n}$, on a quantum channel $q_m$, at wavelength $\lambda_{q_m}$.
For a classical signal of intensity $I$, the expected power of light due to Raman scattering in the forward direction is given by,
\begin{equation}\label{eqn:Raman}
    I^{f} = I \, e^{-\alpha L} \, L \, \Gamma(\lambda_{c_n},\lambda_{q_m}) \, \Delta \lambda,
\end{equation}
where $I$ is the optical intensity of the classical channel,
$\alpha$ is the fibre attenuation coefficient,
$L$ is the optical fibre length,
$\Gamma(\lambda_{c_n},\lambda_{q_m})$ is the Raman cross section (per fibre and bandwidth),
and $\Delta \lambda$ is the bandwidth of the quantum receiver~\cite{Bahrani2016OptimalNetworks}.
The Raman cross section, $\Gamma(\lambda_{c_n},\lambda_{q_m})$, is dependent on the wavelength of the classical channel $\lambda_{c_n}$ and the wavelength of the quantum channel $\lambda_{q_m}$.
To define this value we rewrite the Raman cross section as $\rho(\Delta\lambda)$, where $\Delta\lambda = \lambda_{q_m}-\lambda_{c_n}$ is the difference in wavelength between the classical channel and the quantum channel, as given in Eraerds et al.~\cite{Eraerds_2010}.
This substitution was verified against measurement of background noise in quantum communications from Raman scattering, from Tessinari et al.\cite{tessinari2021towards}.

This calculation can then be used to compare expected levels of Raman scattering, due to unavoidable lengths of SMF between the wavelength filtering of the classical communications channels, and the wavelength demultiplexing of the quantum and classical channels, as shown in \Cref{fig:testbed} \textbf{a)}.
We will also compare the performance of the HCF, in reducing the Raman scattered noise, and that of a similar fibre length of SMF with the same total system loss.

\section{Experimental Results}
\label{sec:result}

In this work we use an entanglement-based QKD protocol to analyse the entanglement links, specifically
using the BBM92-based QKD protocol \cite{bennett1992quantum}, to provide direct measurements of the rate and quality of entanglement distribution.
In this protocol, the noise seen on each link can be clearly examined by comparing the Quantum Bit Error Rate (QBER) of each link.
If the background noise increases, such as if we were to expect scattered light from non-linear scattering \cite{fan2023energy,tessinari2021towards}, the number of accidental coincidences seen in the QKD protocol increases, resulting in an increased QBER.
The rate of entanglement can be directly taken from the SKR of the QKD.

\subsection{3-user network}

\begin{figure*}[!t]
    \centering
    \includegraphics[width=130mm]{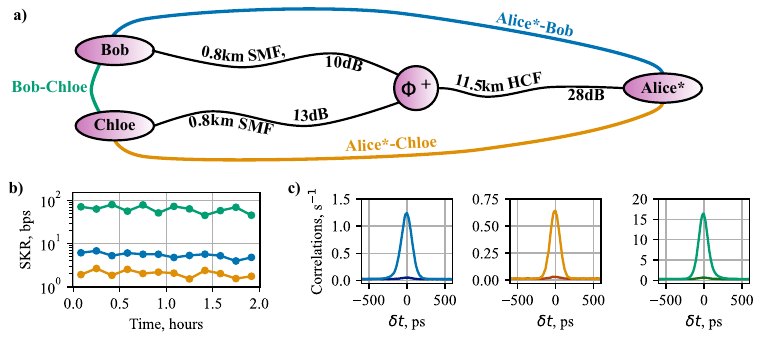}
    \caption{
    The 3-user quantum network with coexistence of classical light through the HCF.
    \textbf{a)} shows the physical network topology with fibre links in black lines, as shown in \Cref{fig:testbed} \textbf{a)}, and the logical entanglement connectivities between the users in coloured lines.
    Link losses include contribution from fibre, q-ROADM, and classical light filtering.
    \textbf{b)} shows the secret key rate (SKR) of each entanglement link over $2$\, hours, where each point is a $10$\,minute average.
    \textbf{c)} shows the correlation histograms of each link, where the lighter peak is the expected correlation results and the darker peak is the erroneous (noise) correlations.
    }
    \label{fig:3U_data}
\end{figure*}

\begin{table}[!b] 
\centering
\caption{
Statistics for the 3-user entanglement network. 
The error on the QBER is poissonian error from the temporal cross-correlation data.
}
\label{tab:3U_results}
\begin{tabular}{ccc}
\hline
\hline \\[-20pt]
\multicolumn{1}{c}{Link}  &   \multicolumn{1}{c}{SKR, bps}  &   \multicolumn{1}{c}{QBER, \%}      \\
\hline \\[-20pt]
Alice*-Bob    &   5.47  &  $5.0\pm0.4$   \\[-4pt]
Alice*-Chloe  &   2.05  &  $5.5\pm0.6$   \\[-4pt]
Bob-Chloe     &   63.0  &  $5.6\pm0.6$   \\
\hline
\hline \\[-10pt]
\end{tabular}
\end{table}

To gather initial data, a 3-user quantum network was constructed, following the testbed structure of \Cref{fig:testbed}.
The physical and logical topologies can be seen in \Cref{fig:3U_data}~\textbf{a)}.
On this network, there are two entanglement links that coexist with the classical communication.
These are the Alice*-Bob link and the Alice*-Chloe link, which experience a combined loss of $38$\,dB and $41$\,dB respectively.
The remaining link, Bob-Chloe, experience no coexistence and has a system loss of $23$\,dB.

\Cref{fig:3U_data}~\textbf{b)} shows the asymptotic  SKR or each link over 2 hours.
Since Alice*’s quantum channels are transmitted through extra components (CPL, HCF, Demux, and Mux from \Cref{fig:testbed}) to facilitate the coexistence, this increases the overall optical loss which reduces the SKR.
Therefore, the Alice*-Bob and Alice*-Chloe links have a high loss, hence leading to a lower SKR of $5.47$\,bps and $2.05$\,bps respectively, as shown in \Cref{tab:3U_results}.
In contrast, with the significantly lower loss of the Bob-Chloe link, a SKR of $63$\,bps is achieved.

The correlation histograms of each link can be seen in \Cref{fig:3U_data}~\textbf{c)}.
If additional single photon counts were seen on the detectors from scattering phenomena in the fibre, then an increased background of accidental detections would be expected. 
As the background in these plots is of a similar order from the peak of the histogram, we can conclude that only a negligible contribution of the overall single photon counts is coming from scattered light.

\subsection{4-user network}

\begin{figure}[!b]
\centering
\includegraphics[width=8.5cm]{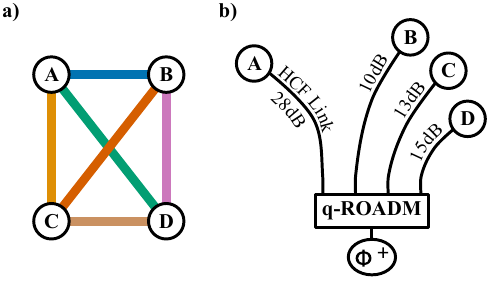}
\caption{
The logical topology of the entanglement links between users, \textbf{a)}, and a sketch of the physical topology of the fibre network, \textbf{b)}, as detailed in \Cref{fig:testbed} where link losses include all optical components between the entanglement source and the quantum measurement module, including loss associated with the q-ROADM.
}
\label{fig:4UTopologies}
\end{figure}

\Cref{fig:4UTopologies} shows network topologies explored in this section.
\Cref{fig:4UTopologies}~\textbf{a)} shows the logical topology of the entanglement links between users in the network.
A full mesh 4-user network has 6 links, requiring 12 wavelengths (6 entangled pairs).
Each user then receives 3 quantum channels, each using a different wavelength sharing entanglement with another wavelength. 
A coloured line in the logical topology defines a quantum link, where one channel in a pair is distributed to one party, and the corresponding channel goes to another party.
These channels are then distributed over fibre links shown in \Cref{fig:4UTopologies}~\textbf{b)}, to produce the logical-topology shown in \Cref{fig:4UTopologies}~\textbf{a)}.

\Cref{fig:4U_summary} shows experimental results of the entanglement network, where the coexisted classical channels to Alice* are switched on and off every hour, testing over $55$\,hours.
Each data point is the average over $1$\,hour of data, Alice* receiving coexisted light is represented by dots, and the linear fit a solid line, and data where no coexisted light is present is shown with a star symbol of a darker tone, and a dashed linear fit.
Links with no coexistence can be seen in \Cref{fig:4U_summary}~\textbf{a)}.
This shows that the linear fits of both sets of data (with coexistence on and off) exhibit the same linear fitting, with a slight downwards trend over the $55$\,hours.
This downward trend in SKR comes from a slow power drift of the pump laser over the course of the $55$\,hours.

Each link with coexisted classical channels shows a reduction in the SKR between the classical channels being turned on and turned off, as shown in \Cref{tab:4U_results} \textbf{b)}-\textbf{d)}.
This SKR drop is caused by an increase in the QBER of $1.5-2$\% per link, due to an extra $\approx 2000$ single-photon counts per second (cps) noise seen on the single photon detectors of Alice*.
The correlation histograms of the three entanglement links with Alice* are shown in \Cref{fig:4U_summary} \textbf{e)}-\textbf{g)}.
The dashed lines show the expected accidental coincidence counts from the single photon detections at the two users.
This increase in the background reduces the visibility of each histogram, as shown in \Cref{tab:4U_results}.
We can also see the stability of the background in a link without coexistence in \Cref{fig:4U_summary} \textbf{h)}.

\begin{table}[!t] 
\centering
\caption{
Statistics for the 4-user entanglement network
The error on the QBER is poissonian error from the temporal cross-correlation data.
}
\label{tab:4U_results}
\begin{tabular}{ccccc}
\hline
\hline \\[-15pt]
\multicolumn{1}{c}{}  &   \multicolumn{2}{c}{Co-existence on}  &   \multicolumn{2}{c}{Co-existence off}     \\[-5pt]
Link          & SKR, bps & QBER        &  SKR, bps & QBER    \\
\hline \\[-15pt]
Alice*-Bob    & 0.70     & $6.6\pm0.4$ & 0.93      &  $5.1\pm0.3$  \\[-4pt]
Alice*-Chloe  & 0.11     & $8.8\pm0.8$ & 0.25      &  $6.7\pm0.6$  \\[-4pt]
Alice*-Dave   & 0.053    & $10\pm0.8$  & 0.20      &  $8.5\pm0.6$  \\[-4pt]
Bob-Chloe     & 9.8      & $6.1\pm0.1$ & 9.5       &  $6.0\pm0.1$  \\[-4pt]
Bob-Dave      & 16       & $7.0\pm0.1$ & 16        &  $6.8\pm0.1$  \\[-4pt]
Chloe-Dave    & 9.8      & $7.3\pm0.1$ & 12        &  $6.6\pm0.1$  \\
\hline
\hline \\
\end{tabular}
\end{table}

In such a system there are 3 main causes of accidental photon detections.
These are, ambient noise counts, Raman scattered light, and cross-talk after wavelength-division filtering.
As we can see the additional counts only when the classical channels are activated and as there is no effect on the background seen in \Cref{fig:4U_summary}~\textbf{h)}, we can eliminate the ambient noise counts component.
Using \Cref{eqn:Raman}, we can calculate the expected Raman contribution for the short lengths of SMF that is present in the link.
Between the CPL and the Demux, shown in \Cref{fig:testbed} \textbf{a)}, there is appropriately $1$\,m of SMF present.
This would lead to $\approx 700$\,cps of Raman scattered photons are the quantum measurement of Alice*.
This leaves some $\approx 1400$\,cps at Alice* beyond the expected Raman noise.
We would expect this to be caused by insufficient wavelength filtering isolation on the Demux and Mux in Alice* in \Cref{fig:testbed}.

Separately, there is a reduction in SKR across all links in the entanglement network, between the 3-user and 4-user demonstrations.
This reduction comes from a reduction in the pumping power of the entanglement source and a small contribution to the polarisation compensation gradually became suboptimal at the time of data collection over 55 hours.
This can be seen by comparing the SKR and QBER values with background subtraction and without such corrections. 
Typically background subtraction is not performed, but for link analysis it can be used to find causes of the increased QBER.
\Cref{tab:QBER_comparison} shows that the QBER values for all links are higher in the 4-user demonstration for both the background subtracted and background included data compared to 3-user network.
It would be expected that the background subtracted QBER would remain the same if the increased QBER was only from the additional photon detections.
Of the three links that are present in both the 3-user and 4-user demonstrations, only the Bob-Chloe link has similar background subtracted QBER.
This suggests that the contributions to the increased QBER do not come from the entanglement source, but instead come from the polarisation neutralisation of the fibre links to Alice and Dave.

Comparing the Bob-Chloe link between the 3-user and 4-user cases we see a reduction of the SKR from $72$\,bps to $9.8$\,bps, showing a $\approx 7$ times reduction in SKR. 
This significant reduction in the SKR would not be seen from only the increased QBER, from $4.1$\% to $4.9$\%, but would come from the reduction in source pumping power with single photon detections from $183$\,kcps to $61$\,kcps per received channel at Bob, and from $106$\,kcps to $40$\,kcps per received channel at Bob.
This $\approx 65$\% reduction in the single photon counts, along with a decreased signal-to-noise ratio, by an additional quantum channel for all users for the 4-user network scenario, would result in a significant reduction in the SKR.
This effect would be expected for all quantum channels in the network, and a reduction in SKR across the network is expected.

\afterpage{
  \clearpage
  
\begin{figure*}[!t]
\centering
\includegraphics[width=130mm]{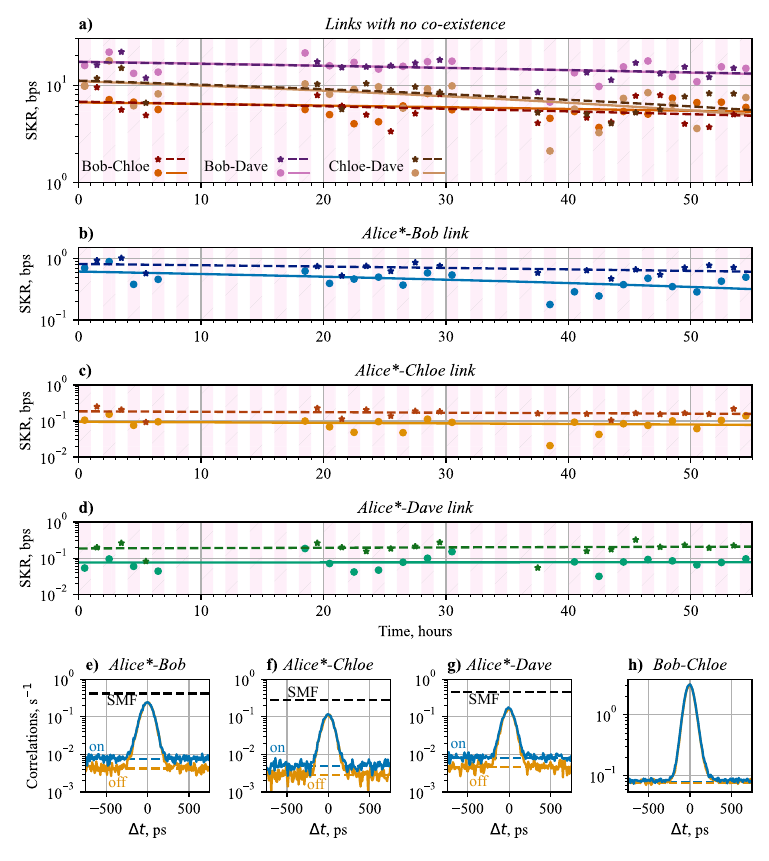}
\caption{
The 4-user network data, with a total of 6 entanglement links.
\textbf{a)}-\textbf{d)} shows the secret key rate (SKR) of the links over a $55$\,hour run.
Dots show when coexisted classical light is present, and stars show when coexisted classical light being turned off.
Solid lines show the linear fit of the coexisted SKR, and dashed lines show the linear fit of the SKR with no coexisted light.
\textbf{a)} shows the SKR of the three links with no-coexisted light, specifically between Bob, Chloe, and Dave.
\textbf{b)}, \textbf{c)}, and \textbf{d)} shows the SKR of the Alice*-Bob, Alice*-Chloe, and Alice*-Dave links respectively.
\textbf{e)}-\textbf{h)} shows example correlation histograms for links in the entanglement network, where the correlation value is given per $10$\,ps bin.
\textcolor{cbl0}{Blue} lines show data from when classical channels are on, and \textcolor{cbl1}{orange} shows when coexistence was off.
Dashed lines shows the accidental correlation background calculated from the single photon detection rates.
The black dashed line shows the accidental correlation rate if the HCF was replaced with a standard SMF fibre of the same length.
}
\label{fig:4U_summary}
\end{figure*}

\clearpage
}

\subsection{The effect of Raman scattering}

Following Sec.~\ref{subsec:Raman} we can analyse the effect of Raman scattered noise, in the situation where the HCF is replaced with a typical SMF fibre\footnote{Corning, Corning® SMF-28® Ultra Optical Fibre (2021). PI-1424, Available at \url{https://www.corning.com/media/worldwide/coc/documents/Fiber/product-information-sheets/PI-1424-AEN.pdf}}, 
with additional loss of the HCF added as a loss after this optical fibre.
Here, we would expect the added noise to increase the total rate of photon detections to 
increased to $\approx 280$\,kcps for the Alice* measurement system.
The expected accidental coincidence rate would then become $0.4$ coincidences per second per bin for the Alice*-Bob link.
This is an increase from $0.0075$ accidental coincidences per second per bin, an increase of 2 orders of magnitude.
Using SMF instead of HCF would immediately prevent any secure key from being generated on this entanglement link, reducing the visibility of the signal from the background from $94.3$\%~($91.2$\%) without~(with) coexisted classical light through HCF to $14.4$\% with coexisted classical light through SMF.
This accidental correlation background can be seen in \Cref{fig:4U_summary} \textbf{e)}-\textbf{g)}.

\begin{table}[!t] 
\centering
\caption{The QBER value for the 3-user and 4-user network experiment with a fixed 100ps bin in the correlation histogram}
\label{tab:QBER_comparison}

\begin{tabular}{ccccc}
\hline
\hline \\[-15pt]
\multicolumn{1}{c}{}  &   \multicolumn{2}{c}{Background included}  &   \multicolumn{2}{c}{Background subtracted}     \\[-3pt]
Link          & Visibility & QBER        &  Visibility & QBER    \\
\hline

\\[-15pt]
\multicolumn{5}{c}{3-user demonstration} \\
\\[-20pt]
Alice*-Bob    & $92.3$\%  & $3.9$\%      & $95.5$\%    &  $2.3$\%  \\[-4pt]
Alice*-Chloe  & $90.6$\%  & $4.7$\%      & $94.8$\%    &  $2.6$\%  \\[-4pt]
Bob-Chloe     & $91.9$\%  & $4.1$\%      & $95.1$\%    &  $2.4$\%  \\

 \\[-20pt]
\multicolumn{5}{c}{4-user demonstration} \\
 \\[-20pt]
Alice*-Bob    & $86.4$\%  & $6.8$\%      & $92.8$\%    &  $3.6$\%  \\[-4pt]
Alice*-Chloe  & $82.7$\%  & $8.6$\%      & $92.2$\%    &  $3.9$\%  \\[-4pt]
Alice*-Dave   & $79.7$\%  & $10.1$\%     & $87.6$\%    &  $6.2$\%  \\[-4pt]
Bob-Chloe     & $90.3$\%  & $4.9$\%      & $95.5$\%    &  $2.2$\%  \\[-4pt]
Bob-Dave      & $87.7$\%  & $6.1$\%      & $91.9$\%    &  $4.1$\%  \\[-4pt]
Chloe-Dave    & $86.8$\%  & $6.6$\%      & $90.4$\%    &  $4.8$\%  \\
\hline
\hline \\
\end{tabular}
\end{table}

\section{Conclusion}
\label{sec:conclusion}

We present the distribution of polarisation-entangled photons over an 11.5~km long HCF in a 4-user fully connected entanglement-based quantum network. 
We also demonstrate a 4-user (3-user) entanglement network with coexistence of three (two) quantum channels and four 200~Gbps classical channels over the same HCF, with a channel separation of $1.6$\,nm. 
The Bell state fidelity of $>80$\%, is sufficient for efficient QKD key generation that we perform between the three users. 
In the 3-user network, the three users had a similar QBER value of $\approx 5.5$\%
for both coexistence (A-B) and (A-C) and no-coexistence (B-C) links, affirming the advantage of HCF for the quantum and classical light coexistence scheme.
When expanding to the 4-user interconnection, the network can still execute BBM92 protocol and generate a secure key.
However, the QBER on all pre-existing links was increased when compared the the 3-user network.
The QBER increases due to a reduced signal-to-noise ratio with a reduction in the pump power of the entanglement source, with only a $\approx 3$\% decrease in background subtracted Bell state fidelity between all links in \Cref{tab:QBER_comparison}.

This shows that entanglement distribution quantum networking is compatible with HCF with many users connected simultaneously with a fully-connected topology. 
Newer generations of the same HCF technology have begun to demonstrate losses lower than available Ultra Low Loss solid core SMF.
This makes HCF the ideal candidate for future quantum networks.
With constant improvements of HCF, entanglement networks can be connected over longer distance with coexistence on all links, allowing for true heterogeneous quantum-classical networking.

\begin{backmatter}

\bmsection{Funding}
This work was funded by EU Horizon 2020 funded project UNIQORN (820474) and the EPSRC Airguide Photonics Collaboration Fund Award (ref: 517129) (EP/P030181/1). Part of the research leading to this work has been supported by the Quantum Communication Hub funded by the EPSRC grant ref. EP/T001011/1 and the ERC LightPipe project (grant n$^{\circ}$ 682724). We also acknowledge the support of EP/Z533208/1 and EP/X039439/1 from EPSRC.

\bmsection{Author Contributions}
OA constructed the coexistence test-bed and characterised the system.
OA and RW constructed the q-ROADM used in the test-bed.
MJC produced the entangled photon pair source, maintained the quantum measurement devices, collected the experimental data, produced control code used in the system, and performed data analysis for the project.
SB assisted in setup of the quantum network.
GTJ, HS, PP, and FP produced the hollow core fibre used in the test-bed.
JR, RN, SKJ, RW, and DS provided supervision of the project.
OA and MJC were major contributors in writing the manuscript.
MJC prepared all figures and tables.
MJC prepared the data and code repository for reader access.
This manuscript was significantly improved by the detailed feedback from SB,  PP, FP, JR, SKJ, and RW. 
All authors read and approved the final manuscript.

\bmsection{Acknowledgment}
We are grateful to Dr John R. Hayes for invaluable assistance in the fabrication of the hollow-core fibres used in this work.

\bmsection{Competing Interests}
All authors declare no financial or non-financial competing interests

\bmsection{Data availability}
Data files and python code are available at the University of Bristol data repository, \url{data.bris.ac.uk}, at \url{https://doi.org/10.5523/bris.61193flfovbn2puknndsr3qm5}.
Raw data files consisting of time tag data may be available on request from the contact author, due to the file sizes of the raw single photon count data sets.
The available code is used to take the csv summary data to produce all plots in this article.
Code is provided that shows how the SKR and QBER of the entanglement links is produced from the correlation histogram data. 
Full data analysis code is not provided, but may be made available on request.


\end{backmatter}



\end{document}